\documentclass[aps,prx,reprint,groupedaddress,amsmath,amssymb,longbibliography]{revtex4-2}

\usepackage{graphicx}
\usepackage{dcolumn}
\usepackage{bm}
\usepackage{xcolor}

\begin{document}

\title{Photoelectron energy peaks shift against the radiation pressure in strong field ionization }

\author{Kang Lin$^{1,2}$}
\email{lin@atom.uni-frankfurt.de}
\author{Sebastian Eckart$^1$}
\email{eckart@atom.uni-frankfurt.de}
\author{Alexander Hartung$^1$}
\author{Daniel Trabert$^1$}
\author{Kilian Fehre$^1$}
\author{Jonas Rist$^1$}
\author{Lothar Ph. H. Schmidt$^1$}
\author{Markus S. Sch\"offler$^1$}
\author{Till Jahnke$^1$}
\author{Maksim Kunitski$^1$}
\author{Reinhard D\"orner$^1$}
\email{doerner@atom.uni-frankfurt.de}

\affiliation{$^1$ Institut f\"ur Kernphysik, Goethe-Universit\"at, Max-von-Laue-Str. 1, 60438 Frankfurt am Main, Germany \\
$^2$ State Key Laboratory of Precision Spectroscopy, East China Normal University, 200241, Shanghai, China 
}

\date{\today}
\begin{abstract}
The photoelectric effect describes the ejection of an electron upon absorption of one or several photons. The kinetic energy of this electron is determined by the photon energy reduced by the binding energy of the electron and, if strong laser fields are involved, by the ponderomotive potential in addition. It has therefore been widely taken for granted that for atoms and molecules the photoelectron energy does not depend on the electron's emission direction but theoretical studies have questioned this since 1990. Here we provide experimental evidence, that the energies of photoelectrons emitted against the light-propagation direction are shifted towards higher values while those electrons that are emitted along the light-propagation direction are shifted to lower values. We attribute the energy shift to a nondipole contribution from the interaction of the moving electrons with the incident photons.
\end{abstract}

\maketitle

\section{Introduction}
In 1905, the concept of light quanta (termed as photons nowadays) was first proposed by Einstein to explain the photoelectric effect, where a bound electron can only be released to the continuum by absorbing a single photon of an energy which is larger than the ionization potential \cite{1_einstein1905erzeugung}. The final kinetic energy of the liberated electron is $E_e = \hbar\omega - I_p$, where $\omega$ is the light's frequency and $I_p$ is the ionization potential of the target. Later, in 1931, Goeppert-Mayer \cite{9_GoeppertMayer1931} showed that bound electrons can absorb multiple photons simultaneously such that the electron energy is given by $E_e = n\hbar\omega - I_p$, where $n$ is an integer that is large enough such that the bound electrons can overcome $I_p$. In 1979, with the advent of pulsed laser techniques, P. Agositni \textit{et al.} found that even free-free transition can happen if the laser intensity is high enough \cite{10_agostini1979p}, or – more generally speaking – a bound electron can absorb more photons than needed to overcome $I_p$. As a result, a series of discrete peaks in the photoelectron's energy domain occur, which are equally spaced by the photon energy of the driving laser field. The phenomenon has been termed above threshold ionization (ATI) \cite{11_Freeman1987,6_Keldysh1965,7_Perelomov1966,8_Milosevic2006} and is shown schematically in Fig. \ref{fig1}. 

In case of strong field ionization, the electric field strength of the driving laser pulse is comparable to the atomic or molecular Coulomb field. Thus, the field-free ionization potential of the atom or molecule is Stark-shifted \cite{12_Shakeshaft}, leading to an effective potential $I_p+U_p$ with the ponderomotive energy $U_p$ that is given in atomic units by \cite{11_Freeman1987}:
\begin{equation}
\label{Eq1}
U_p = \frac{F^2}{4\omega^2}
\end{equation}
Here, $F$ is the peak electric field of the linearly polarized driving field. In principle, the increased effective ionization potential can be compensated by the gradient force in the focus if the laser pulse is long enough to allow the electron to escape the focus before the laser pulse has faded out. However, for ultrashort laser pulses, with durations of tens of femtoseconds, this is not the case and thus the final kinetic energy of photoelectrons produced from ATI process is given by \cite{11_Freeman1987}: 

\begin{equation}
\label{Eq2}
E_\mathrm{ATI,n}=n\hbar\omega - I_p - U_p
\end{equation}

Already in 1990, Reiss suggested that $U_p$ has to be replaced by an effective ponderomotive potential \cite{2_reiss1990relativistic} leading to Eq. (\ref{Eq3}) which spawned further theoretical work by Böning \textit{et al.}, Jensen \textit{et al.}, Lund and Madsen, as well as Brennecke and Lein. \cite{3_Boning2019,4_Jensen,5_Lund2021,Brennecke2021_PRA}

\begin{equation}
\label{Eq3}
U_p^\mathrm{eff} (p_x)=U_p/(1-p_x/c)\approx (1+p_x /c)U_p
\end{equation}

This expression for $U_p^\mathrm{eff}$ includes nondipole \cite{13_Ludwig2014,14_Ivanov2016,15_connerade2000controlled,16_Chelkowski,17_Chelkowski2014,18_Brennecke2018,19_Danek_2018,20_Smeenk2011_B} contributions to the light-matter interaction which depend on $p_x$ (see Eq. (21) in Ref. \cite{3_Boning2019}). Here, $p_x$ is the electron momentum along the light-propagation direction. Accordingly, this prediction suggests that ATI energy peaks do not generally occur as a series of equally spaced peaks, but that the position of the n$^\mathrm{th}$ ATI peak depends on the electron momentum in the light-propagation direction $p_x$:

\begin{equation}
\label{Eq4}
\tilde{E}_\mathrm{ATI,n} (p_x)=n\hbar\omega - I_p - (1+p_x /c)U_p  
\end{equation}

\onecolumngrid\
\begin{center}\
\begin{figure}[h]\
\includegraphics[width=0.9\columnwidth]{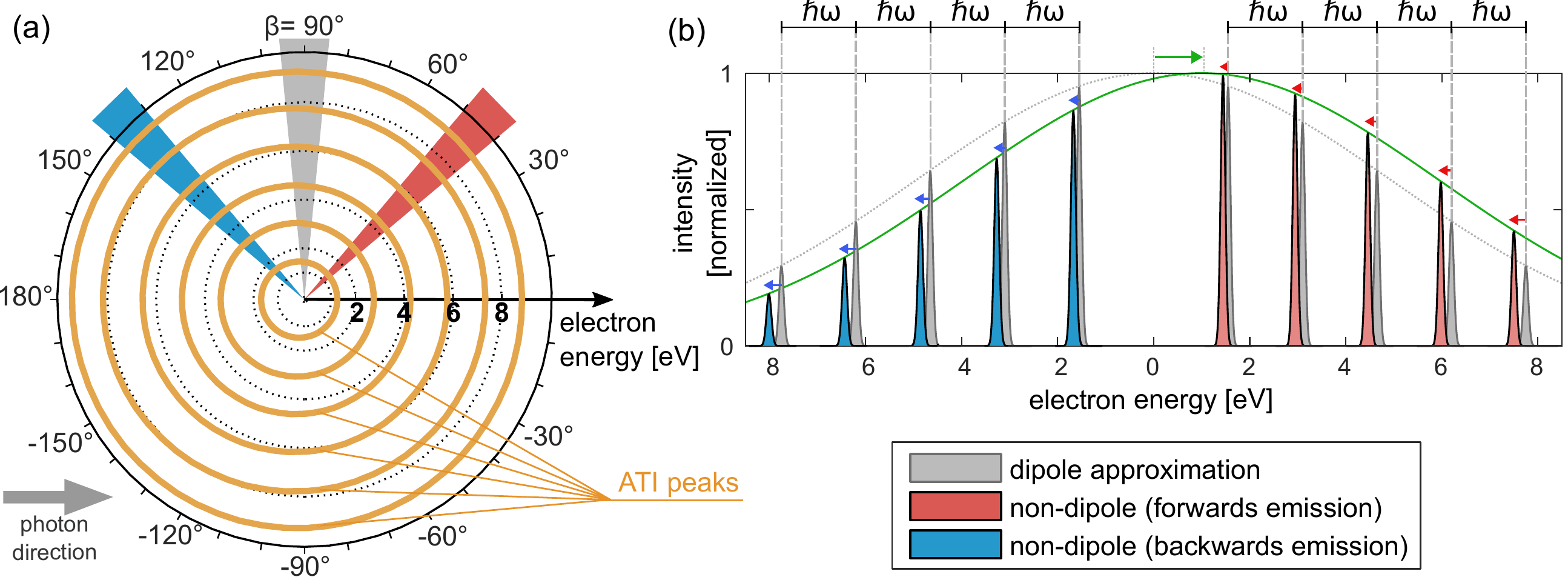}
\caption{Illustration of ATI beyond the dipole approximation. (a), Within the dipole approximation, a series of concentric circles is formed in energy space, which corresponds to peaks in the energy distribution that are equally spaced by the photon energy. The angle $\beta$ indicates the electron emission direction with respect to the light-propagation direction. For the nondipole case, the concentric circles are deformed (yellow curves, Eq. (\ref{Eq4})). The deformation is exaggerated for visualization purposes by setting the speed of light to $c=0.2\cdot 137$\,a.u.$=27.4$\,a.u. in Eq. (\ref{Eq4}). The gray dashed lines are circles to guide the eye. (b), The energy spectra for the subsets indicated in (a). The colors in (b) correspond to the colors in (a). It is clearly seen that the spacing of the ATI energy peaks, which is often considered to be equal to the photon energy $\hbar\omega$, depends on the electron's emission direction and that the ATI peaks are shifted backwards (blue and red arrows). The green line and the green arrow illustrate the shift of the envelope of the ATI peaks (not to scale) that is due to the photon's momentum and which is opposite to the shift of the ATI peaks.}
\label{fig1}
\end{figure}\
\end{center}\
\twocolumngrid\

Figure 1 illustrates the effect of this nondipole correction on the photoelectron spectrum and highlights that for certain emission angles the ATI peaks are not equally spaced anymore as depicted in Fig. \ref{fig1}(b). Thus, the difference $\tilde{E}_\mathrm{ATI,n} (p_x)-E_\mathrm{ATI,n}$ increases with the electron momentum in the light-propagation direction $p_x$. The nondipole shift points for all ATI peaks into the direction that is opposite to the radiation pressure. However, the envelope (green line in Fig. \ref{fig1}(b)) is shifted in the direction that points along the radiation pressure). It should be noted that previously, an appearingly similar backwards shift of the envlope for very low electron energies has been observed by Ludwig \textit{et al.} Ref. \cite{13_Ludwig2014}. Importantly, the shift that we observe is present for all ATI peaks and even vanishes for very low energy electrons and is thus very different to the one reported by Ludwig \textit{et al.} What is the reason for the backwards shift of the ATI peaks?

\section{A Doppler-like Effect}
An intuitive understanding of the nondipole energy shift can be gained using a simple perspective based on a Doppler-like effect \cite{3_Boning2019}. By analogy with the Doppler shift for a moving object, the frequency of the laser field's time-dependent force acting upon the electron in the laboratory frame is given by $\omega_\mathrm{eff}=\omega(1-p_x /c)$ in atomic units. Thus, this Doppler-like effect would occur also in the absence of the magnetic field of the incident light. Inserting $\omega_\mathrm{eff}$ into Eq. (\ref{Eq1}) results in a predicted energy for the n$^\mathrm{th}$ ATI peak:

\begin{align}
\label{Eq5}
\tilde{E}_\mathrm{ATI,n}^\mathrm{doppler} (p_x)= n\hbar \omega - I_p - \frac{U_p}{(1-p_x /c)^2} \nonumber \\
\approx n\hbar \omega - I_p- (1+2p_x/c)U_p
\end{align}

Equation (\ref{Eq5}) follows the intuitive picture that is described above but surprisingly Eq. (\ref{Eq4}) and Eq. (\ref{Eq5}) differ by a factor of two in front of $p_x/c$.

The energy shift expected from  Eq. (\ref{Eq4}) is extremely small for typical experimental conditions in the strong field regime. For example, a linearly polarized laser pulse with a central wavelength of 800\,nm and a peak intensity of $1.0 \times 10^{14}$ W/cm$^2$ has a ponderomotive potential of 6.0 eV. For an initial momentum component of $p_x=0.2$\,a.u. the expected change in ATI peak energy is only $U_p p_x /c\approx 9$\,meV.

\section{Experiment with Counter-Propagating Laser Beams}
The key to successfully resolve such a small energy shift is to minimize systematic errors by using an experimental setup that allows for the ionization of individual atoms or molecules from a gas jet in an experimental geometry where the light-propagation direction can be inverted while everything else remains unchanged. To this end, we employ the same experimental setup as in Refs. \cite{21_Hartung2021,22_hartung2019magnetic}. Briefly, two counter-propagating laser beams are focused onto the same spot in a gas jet to trigger the ionization process. The energy spectra are recorded by a specialized cold target recoil ion momentum spectroscopy (COLTRIMS) setup with extremely high momentum resolution in light-propagation direction. With this setup it is possible to record energy spectra under the exact same experimental conditions except for the inversion of the light-propagation direction by employing two different laser pathways.  

The two counter-propagating laser beams were generated from a Ti:Sapphire laser system (Coherent Legend Elite). The output of the laser system (25fs, 800nm, 10kHz) is split into two pathways using a dielectric beam splitter. The intensity and polarization of each laser pathway can be adjusted independently. Eventually, the two linearly polarized laser beams are focused into the vacuum chamber of a COLTRIMS reaction microscope \cite{35_ullrich2003recoil} from two opposite sides using two independent lenses (f=25cm) onto the same spot inside a supersonic gas jet of H$_2$ molecules. For both laser pulses the polarization axis is aligned along the z-direction. Two motorized shutters placed in the two beam pathways are used to toggle between both pathways every 3 minutes to minimize systematic errors. The peak intensity in the laser focus is found to be $1.0\times 10^{14}$\,W/cm$^2$ with an uncertainty of $\pm20\%$. For laser intensity calibration, the ratio between double and single ionization yield of xenon atom is recorded and compared to values given in Ref. \cite{36_Chaloupka_2003}. A static electric field of 29.8 V/cm was applied to guide the electrons and ions created from single ionization of H$_2$ molecules to two time- and position-sensitive detectors at opposite ends of the spectrometer \cite{37_jagutzki2002multiple}. The three-dimensional momenta of the electrons and ions were retrieved coincidently from the times-of-flight and positions of impact. The z-direction is the time-of-flight direction of the COLTRIMS reaction microscope. The single event momentum resolution of our COLTRIMS reaction microscope for the detection of a single electron is 0.003 a.u. in $p_x$- and $p_y$- direction and 0.03 a.u. in $p_z$-direction.

\begin{figure}[h]
\includegraphics[width=\columnwidth]{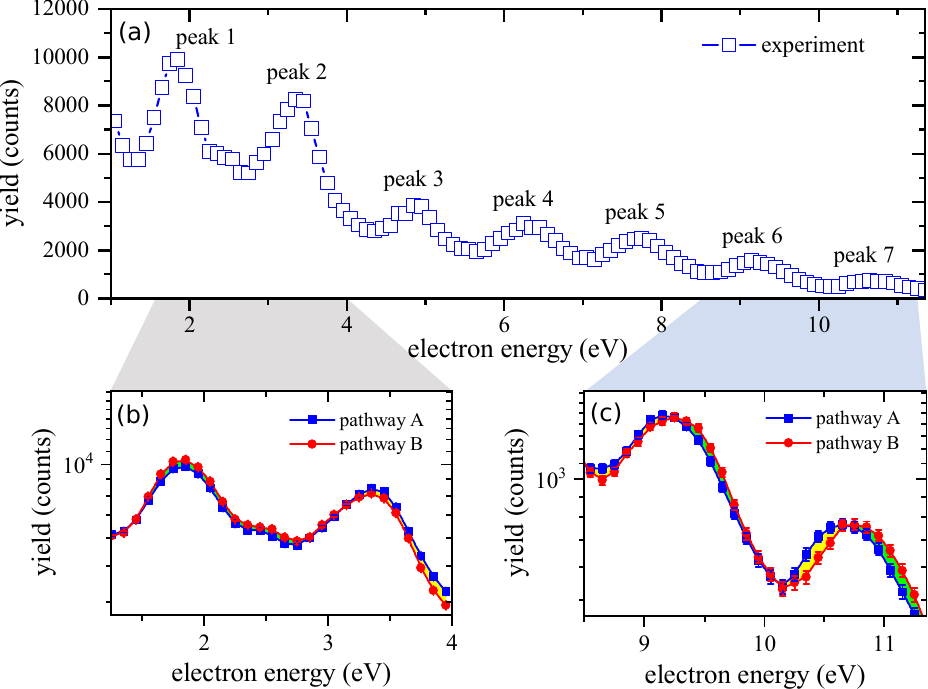}
\caption{Measured electron energy spectra. (a) Full range energy spectrum measured for one of the two counter-propagating laser pathways. (b) and (c) show subsets of (a), where the spectra measured from the two opposite pathways are overlapped. Only the subsets of electrons emitted under an angle of about 10$^\circ$ with respect to the plane perpendicular to the light-propagation direction are considered. The error bars show statistical errors. For panels (a) and (b), the error bars are smaller than the data points.}
\label{fig2}
\end{figure}

Figure 2(a) displays the measured electron energy spectrum for one of the two laser pathways after restricting the data to a subset of electron emission angles of 9$^\circ$ to 11$^\circ$ with respect to the plane that is perpendicular to the light-propagation direction. This corresponds to values of $\beta$ from 79$^\circ$ to 81$^\circ$ and from 99$^\circ$ to 101$^\circ$ in Fig. \ref{fig1}(a). This angular restriction results from a compromise between high statistics for small values of $|p_x|$  and an increased nondipole effect for large values of $|p_x|$. As expected, the two energy spectra collected from the two counter-propagating laser beams are almost identical since the maximum shift is expected to be on the order of only $18$\,meV for $p_x=\pm 0.2$\,a.u. For the low-energy peaks in Fig. \ref{fig2}(b), the peak positions are almost exactly congruent for the two laser beam pathways. This is expected since for electrons emitted along a fixed angle, low electron energies correspond to small values of $p_x$. Strikingly, for the high-energy part of the spectrum, as shown in Fig. \ref{fig2}(c), a clear shift of the peak positions is observed. In addition, the magnitude of the difference, which is on the order of tens of milli-electron volts, agrees with the expectation.

\begin{figure}[h]
\includegraphics[width=\columnwidth]{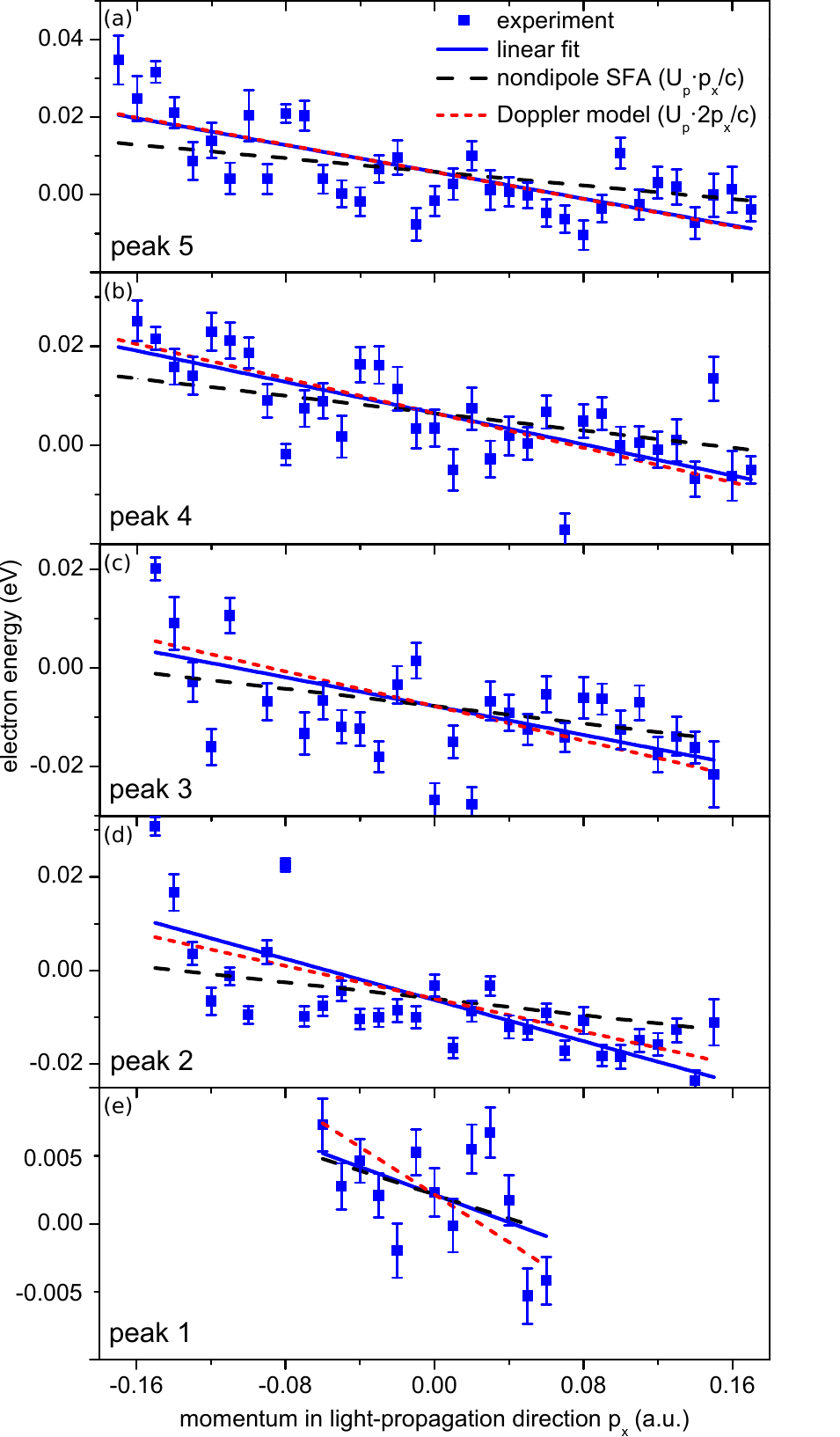}
\caption{Measured nondipole electron energy shift. (a)-(e), Shift of the peak position of the ATI peaks as a function of the momentum in light-propagation direction $p_x$ for the first five ATI peaks (see Fig. \ref{fig2}(a)). The data points are determined by fitting Gaussians to the data and taking the average of the two counter-propagating laser pathways as reference. The expectations from the Doppler model and nondipole strong field approximation (SFA) are shown in red short-dashed and black dashed lines respectively. The error bars show fitting errors.}
\label{fig3}
\end{figure}

\section{Quantitative Analysis}
For a full quantitative analysis, we plot the extracted nondipole energy shift for the first five ATI peaks as a function of $p_x$ in Fig. \ref{fig3}. The position of each ATI peak is obtained by Gaussian fits. The systematic error is minimized by taking the average of the two peak energy values, that are obtained from the two incident beams propagating in opposite directions, as reference. Figure 3 shows the case where the light-propagation direction points to the positive $p_x$-direction. The decrease of the ATI peak energy as a function of $p_x$ is clearly visible. Thus, our experimental results confirm the prediction that the ATI peak energies for electron emission directions that are parallel to the light-propagation direction are lower than for electrons that are emitted anti-parallel to the light-propagation direction. The shift is thus indeed opposite to what one might have naively expected from the direction of the radiation pressure.

\section{Discussion}
\begin{figure}[h]
\includegraphics[width=\columnwidth]{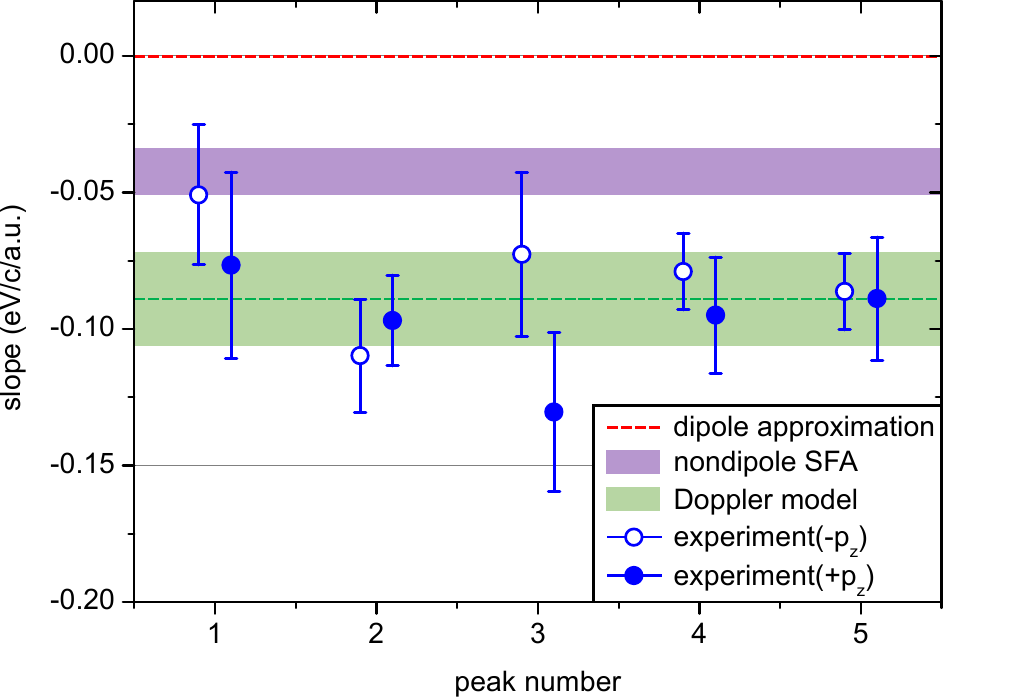}
\caption{Comparison of experimentally obtained nondipole energy shifts and the theoretical predictions. The slopes of the linear fits from Fig. \ref{fig3} are shown in units of eV/c/a.u. (blue open circles) where only negative momenta in z-direction were used (z-axis is the polarization axis). As a cross-check, the filled circles show the same as the open circles but use the data that has positive momenta in z-direction. The red dashed line indicates the slope within the dipole approximation which is zero. The purple and green shaded areas indicate the expectations from the nondipole SFA (Eq. (\ref{Eq4})) and the Doppler model (Eq. (\ref{Eq5})), respectively, by considering an uncertainty of $\pm 20\%$ for the intensity calibration. The error bars show the statistical error.}
\label{fig4}
\end{figure}

Figure \ref{fig4} summarizes the experimental results by depicting the slopes of the linear fits from Fig. \ref{fig3}  for each energy peak separately (see open circles in Fig. \ref{fig4}). In order to compare our experimental findings to the two existing predictions, we show the results together with the expectations from the nondipole strong field approximation \cite{3_Boning2019,4_Jensen} (SFA, Eq. (\ref{Eq4})) and the Doppler model (Eq. (\ref{Eq5})). For the two models we use a ponderomotive potential that corresponds to the laser intensity used in the experiment (taking the uncertainty of the intensity calibration into account). It is surprising that our measurement shows better agreement with the intuitive Doppler model than with the result from the nondipole SFA. 

To further cross-check our experimental findings, we make use of the symmetry of linearly polarized light. For Fig. \ref{fig3} we only analyzed the half of the electron momentum distribution that has negative values for the momentum in the z-direction. The filled circles in Fig. \ref{fig4} are obtained in full analogy to the open circles but for these data points the other half of the electron momentum distribution is used, which has positive momenta in z-direction.

As expected, the filled and open circles show consistent results and indicate that the systematic errors are comparable to the statistical errors. Possible reasons for the experimentally observed enhancement of the energy shift compared to the nondipole SFA model are recollision dynamics \cite{13_Ludwig2014,23_Keil_2017} and Coulomb interaction during and after tunneling \cite{24_Yan2012} which are not included in the nondipole SFA model but could be relevant (especially for linearly polarized light). Further, the nondipole SFA model was developed for atoms. In our experiment we study H$_2$ which is expected to show very similar tunneling dynamics in comparison with atoms \cite{25_Liu_2016}. However, recollisions dynamics are sensitive to the excited states of the ion \cite{26_Wassaf_2004}, which might be a reason for the experimentally observed enhancement of the energy shift compared to the nondipole SFA. In the strong field regime, ATI peaks can be explained by intercycle interference \cite{27_Arbo_2006}. Consequently, different types of interference could contribute to the observed energy shift. For instance, it is conceivable that the interplay of Coulomb interaction with sub-cycle interference \cite{28_Eckart2018SubCycle} or holography \cite{29_Meckel2008,30_Carla2020,31_Huismans2011} is affected by nondipole corrections which could modulate \cite{32_Eckart_2020PRA} the ATI peak positions. This is in line with the findings by Brennecke and Lein who showed that a rigorous treatment of nondipole effects is necessary to model interference using semi-classical models \cite{33_Brennecke}. Moreover, the liberated electron is from a $\sigma$-orbital and thus cannot possess angular momentum. The symmetry of linearly polarized light implies that a potential fingerprint of a non-vanishing angular momentum would not manifest as rotation as in Ref. \cite{34_Kunlong2018}. Future theoretical studies might also investigate the role of properties of the initial electronic state (as e.g. the electron's magnetic quantum number) on the ATI peak positions.

In conclusion, we show that ATI peaks occurring in strong field ionization are only equally spaced in energy for electrons that are emitted perfectly at right angle to the light-propagation direction confirming a theoretic prediction from 1990 \cite{2_reiss1990relativistic}. For all other emission directions, the spatiotemporal evolution of the electric field results in ATI peaks that are not equally spaced in energy. Moreover, we have observed that the nondipole energy shift depends on the x-component of the photoelectron momentum. The ATI peak energies decrease (increase) compared to the expectations from the dipole approximation for electrons that are emitted in the forward (backward) hemisphere. Hence, our results show that the shifts of the ATI peak energies have a direction that is opposite to the direction of the shift that is due to the photon momentum (i.e. light pressure) \cite{21_Hartung2021}. Our findings illustrate that energy conservation is not enough to explain ATI peak energies beyond the dipole approximation and that the spacing of the ATI peaks, which is given by the photon energy in the dipole approximation, depends on the electron's emission direction. We expect that similar corrections should lead to a broadening of the photon's energy spectra for high harmonic generation.

\section{Acknowledgments}
\normalsize
The experimental work was supported by the DFG (German Research Foundation). K. L. acknowledges support by the Alexander von Humboldt Foundation. S. E. acknowledges funding of the DFG through Priority Programme SPP 1840 QUTIF. A. H. and K. F. acknowledge support by the German Academic Scholarship Foundation. We acknowledge helpful discussion with Simon Brennecke, Manfred Lein and Lars Bojer Madsen.

K.L and S.E. contributed equally to this work.

\end{document}